\documentclass{article}

\usepackage{arxiv}

\usepackage[utf8]{inputenc} 
\usepackage[T1]{fontenc}    
\usepackage{hyperref}       
\usepackage{url}            
\usepackage{booktabs}       
\usepackage{amsfonts}       
\usepackage{nicefrac}       
\usepackage{microtype}      
\usepackage{lipsum}

\usepackage[english]{babel}
\usepackage{graphicx}
\usepackage{multirow}
\usepackage{subfigure}

\title{Attention-Based Self-Supervised Feature Learning for Security Data}

\author{
  I-Ta Lee \\
  Purdue University\\
  West Lafayette, IN \\
  \texttt{lee2226@purdue.edu} \\
   \And
  Manish Marwah \\
  Micro Focus Inc.\\
  Sunnyvale, CA \\
  \texttt{manish.marwah@microfocus.com} \\
   \And
 Martin Arlitt \\
  Micro Focus Inc.\\
  Calgary, Canada \\
  \texttt{martin.arlitt@microfocus.com} 
}

\begin{document}
\maketitle

\begin{abstract}
  While applications of machine learning in cyber-security have grown
  rapidly, most models use manually constructed features. This manual approach is error-prone and 
  requires domain expertise. In this paper, we
  design a self-supervised sequence-to-sequence model with attention to learn an embedding for data routinely used in cyber-security applications. The method is validated on two real world public data
  sets. The learned features are used in an anomaly detection model
  and perform better than learned features from baseline methods.
\end{abstract}


\section{Introduction}
Increasingly, enterprises are storing and processing vast amounts of
data to detect security threats and compromises. To build models for
efficient detection, machine learning tools currently show the most promise.
One of the important tasks in building a machine learning model is
coming up with a good set of features. This usually requires deep
knowledge and understanding of the problem domain, and even then
constructing good features for a particular task involves quite a bit
of trial and error.  High quality features constructed for a specific
task (e.g., anomaly detection) simplify model building, and result in
even simple machine learning methods performing sufficiently well.

In recent years, the fields of computer vision and natural language
processing have made great strides using deep learning models to {\em
  learn features} from raw inputs instead of a domain expert manually
engineering features. This has been mainly possible through use of
supervised learning on large scale labelled image data sets (most notably
imagenet \cite{imagenet_cvpr09}) in the case of computer vision, and self-supervised learning on large corpora of texts in the case of natural language processing.

In this paper, we explore the use of feature learning, in place of feature
engineering, for building machine learning models on security
data. Specifically, we look at how the learned features or
representations perform on the popular task of anomaly detection to
detect unusual behavior, which may indicate an attack, a compromise
or a misconfiguration (which may make a system more vulnerable to compromise).
Security data refers to network or host data that enterprises
typically monitor for security threats and attacks. 

We focus on two main challenges faced while learning 
features for security data. First,
security data is unlabelled. Thus, it is difficult to learn features and the
final task jointly as is done in image recognition. Furthermore,
security attacks are ever evolving, so even if a lot of effort is spent
on labeling data, it will only work for known attacks, not for
previously unseen or ``zero-day'' attacks. Secondly, while humans are
good at perceiving and understanding images and natural language, and
can immediately recognize if meaningful features are being learned for these
two modalities, the same cannot be said for say logs of network traffic or
system calls. 

To address the first challenge, we use self-supervised learning, a subset of unsupervised learning, where implicit labels available within the data are used for supervision on a pretext task\cite{jing2019}. In particular, we
use a sequence-to-sequence prediction task, modeled using a 
recurrent neural network model, aided by an attention mechanism. We refer to this method as AS2S (attention-based sequence-to-sequence model). A general solution to the
second challenge is difficult; we explore it in a limited fashion in the context of anomaly detection by using a partially labelled
data set\footnote{The labels are only used for evaluating performance, not for training.} to compute precision and recall. We also visually inspect how the learned features in a two-dimensional embedded space evolve during training. 



To demonstrate the effectiveness of the learned representations, we
use two real data sets to evaluate the performance for anomaly
detection. These data sets include partial ground truth labels
enabling quantification of the results. One of the data sets consists
of network traffic, while the other contains network and host
data. Both of these can be considered as a multivariate temporal
sequence of attributes. The results show that the attention-based
sequence-to-sequence model performs better than the two selected baselines, 
namely a principal component analysis (PCA)-based model and an auto-encoder based
model. Furthermore, we observe as training progresses, the features
learned by our model are better able to separate normal and anomalous data points. 


We make the following key contributions:
\begin{itemize}
    \item We demonstrate how to perform feature learning for security data using attention and self-supervision
    \item We provide performance results for this method on two real world public data sets
\end{itemize}

The rest of the paper is organized as follows.
We discuss related work in the next section, followed by our
solution approach in Section \ref{sec:solution}. Experimental results
are presented in Section \ref{sec:experiments}, followed by
conclusions and future work.

\section{Related Work}

Automatically extracting or learning suitable features from raw data
without human intervention has been a grand challenge in machine
learning.  In fact, the basic idea can be traced back to Herbert Simon
who said in 1969, “Solving a problem simply means
representing it so as to make the solution transparent" \cite{simon1969sciences}.  Over time
diverse techniques \cite{pmlr-v44-storcheus2015survey, jing2019} have
originated in different research communities and include: feature
selection, dimensionality reduction, manifold learning, distance
metric learning and representation learning. Extracting features via
layers of a neural network is commonly referred to as representation
learning \cite{pmlr-v44-storcheus2015survey, bengio2013}, also known as feature
learning and is the focus of this paper in the context of security
data, and its application to machine learning models to detect
security anomalies. 


Anomaly detection in the area of cyber-security has received
considerable research interest and has a long history, although most
deployed systems are rule-based, and application of machine learning
for anomaly detection in security has unique challenges
\cite{2010outside}. Early work tended to use statistics like moving
average or PCA to locate anomalies \cite{2003anoval}. Such methods
worked well in simple cases, but as modern IT systems and data get
larger and more complex, the methods failed to scale with them.
Clustering-based and nearest-neighbor-based methods were proposed
\cite{Scarth1995DetectionON,2007incremental} to utilize locality of
data, but they suffer from both a lack of scalability and the curse of
dimensionality.  

Supervised learning is not attractive for anomaly detection since
labelled data is scarce. Furthermore, especially for security
anomalies, the presence of adversaries make the anomalies (threats and
attacks)  dynamic and constantly changing. Thus, the most common
setting for anomaly detection is unsupervised or self-supervised
\cite{jing2019}. 

Recently, neural-network-based methods have gained much attention, as
the computational performance problem is gradually mitigated by
advanced optimization algorithms and running parallel cores. Xu et
al. \cite{DBLP:journals/corr/XuRYSS15} proposed a de-noising
auto-encoder-based method, which is a prototypical two-phased
framework. An unsupervised learning model, the auto-encoder, is first
trained with a large data set to get the distributional
representations, and later the representations are used in another
domain-related application. However, auto-encoder-based models failed
to facilitate temporal relationships, which is considered an important
component of detecting anomalies. Tuor et al. \cite{tuor2017deep}
adapted a multi-layer LSTM model but with a different special
assumption that the corresponding inputs and outputs of the network
should not differ by more than a threshold; otherwise, it is an anomaly. Similarly,
Malhotra et al. \cite{DBLP:journals/corr/MalhotraRAVAS16} proposed an
LSTM-based encoder-decoder for anomaly detection on machine
signals. Such a model is essentially an auto-encoder between a pair of
sequences, which models temporal information. It can identify
anomalous sequences when the reconstruction error is large. All of
these studies use the underlying principle that fitting models to data
by minimizing reconstruction error can help identify anomalies.  

Attention mechanism is a breakthrough component proposed to
strengthen the encoder-decoder model. In addition to the ability of
decoding quality sequences, it provides soft alignments between the
inputs and outputs, which helps explain the sequence generated. 
More recently BERT/Transformer architecture \cite{bert} was proposed that  uses self-attention. 
Attention mechanism has proved to be useful in neural machine translation
\cite{2015neural,DBLP:journals/corr/BritzGLL17}, machine reading
comprehension \cite{DBLP:journals/corr/HermannKGEKSB15}, and video
captioning \cite{DBLP:journals/corr/XuBKCCSZB15}. 
However, its application to cyber-security anomaly detection has not been explored before. We do not use self-attention, but an auto-encoder with a basic attention component to train the representations.  

In our work, we also adopt the prototypical two-phase learning
framework, as it offers us an opportunity to generalize the use of the
learned representations, which is a solid foundation for building a
robust universal anomaly detection system. An LSTM encoder-decoder
network coupled with attention mechanism is explored in the first
phase of learning universal representations for security
data. Evaluation and analysis on applying the learned representations
in anomaly detection are also investigated.

\section{Our Solution Approach} \label{sec:solution}

\begin{figure}
  \centering
  \includegraphics[width=0.9\textwidth]{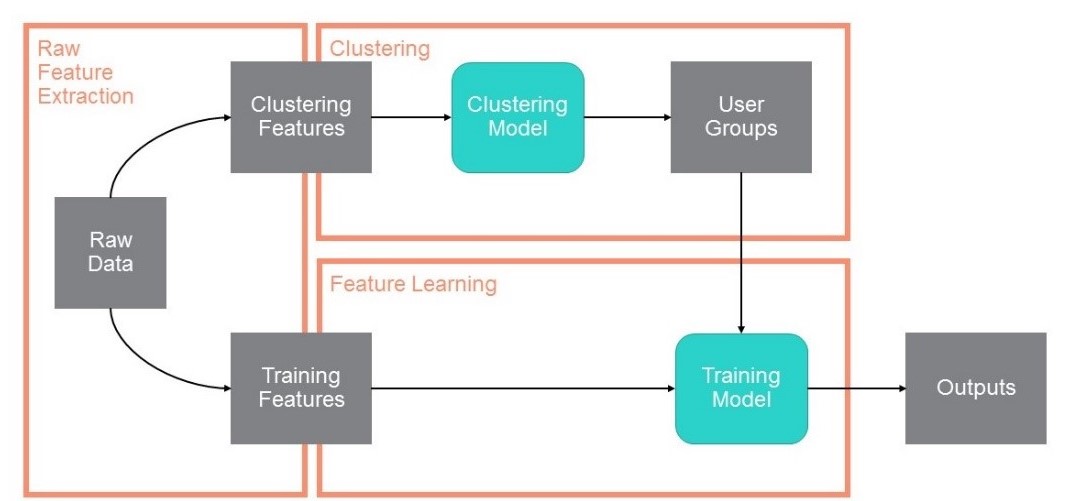}
  \caption{FL4S Framework}
    \label{fig:framework}
\end{figure}

Fig. \ref{fig:framework}. shows the architecture of our approach, which we call  Feature Learning for Security (FL4S) framework. It consists of three main components: Raw Feature Extraction, Clustering, and Feature Learning, which are executed sequentially. The following subsections introduce the data sets used and the three components in detail. 

\subsection{Raw Feature Extraction}

In this initial step, we extract simple features that do not require input from a domain expert. The goal is to use basic representations of the data so  we can perform feature learning on top of them. Security data
can be broadly categorized as network data and host data. 

Network or network traffic data refers to summary information about network communication typically collected at a router or a special appliance installed
in an enterprise network to collect such data. Commonly, such data is collected at an edge router where it captures both ingress and egress
packet data.  A popular network traffic data format is {\em netflow} that includes fields such as IP addresses, port numbers, number of bytes, packets exchanged, application level protocol (inferred from
port information), etc.  Host data refers to data collected at a particular machine, and may include data pertaining to filesystem access,system calls, login and logout information, connection and disconnection of devices, etc.
\begin{figure}
  \centering
  \includegraphics[width=0.90\textwidth]{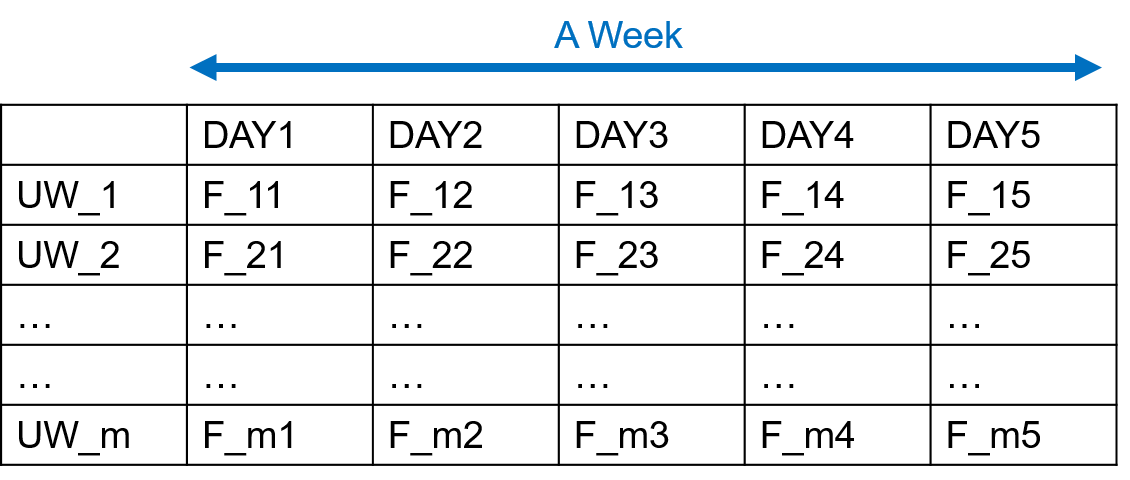}
  \caption{Extracted Raw Features}
    \label{fig:raw_features}
\end{figure}
Security analysts are typically interested in which users or hosts are compromised, or pose a risk. Thus we extract data specific to users.
Further, data generated by users can vary highly over time, and features are aggregated by time windows.  We extract the raw features for each
user and for each pre-defined time window duration. We tried window sizes of 3, 6, 12, and 24 hours; we observed that 24 hour windows gave the best results. Thus, in the following, all references to windows
relate to window sizes of 24 hours. Note that multiple time windows could also be used simultaneously to train multiple models.  
In addition, since users usually have similar data patterns on a weekly basis, we consider
week-long user sequences. Therefore, we redefine each user example as a user-week example. Fig. \ref{fig:raw_features} illustrates this
idea. We also ignore windows during weekends, because they usually have different data patterns than weekdays. As a result, each user-week example has five feature vectors, one for each window
(day). This organization also avoids long sequences for each example, which increases the difficulty of capturing data patterns. Similar
models can be built with weekend data as well.  As shown in Fig. \ref{fig:raw_features}, the data takes the form of a three dimensional tensor: (user-window (UW), time-window, feature ($F_{ij}$)), where each of the $F_{ij}$ is a feature vector.

Security analysts are typically interested in which users or hosts are
compromised, or pose a risk. Thus we extract data specific to users.
Further, data generated by users can vary highly over time, and features
are aggregated by time windows.  We extract the raw features for each
user and for each pre-defined time window duration. We tried window
sizes of 3, 6, 12, and 24 hours; we observed that 24 hour windows gave
the best results. Thus, in the following, all references to windows
relate to window sizes of 24 hours. Note that multiple time
windows could also be used simultaneously to train multiple models.  
In addition, since users
usually have similar data patterns on a weekly basis, we consider
week-long user sequences. Therefore, we redefine each user example as
a user-week example. Fig. \ref{fig:raw_features} illustrates this
idea. We also ignore windows during weekends, because they usually
have different data patterns than weekdays. As a result, each
user-week example has five feature vectors, one for each window
(day). This organization also avoids long sequences for each example,
which increases the difficulty of capturing data patterns. Similar
models can be built with weekend data as well.  As shown in
Fig. \ref{fig:raw_features}, the data takes the form of a three
dimensional tensor: (user-window (UW), time-window, feature ($F_{ij}$)),
where each of the $F_{ij}$ is a feature vector.

\subsection{Segmentation}

A separate model could be built for each user.
However, the number of IP addresses is
typically large, and several IPs may not have enough data associated
with them to train a model. On the other hand, user behavior shows
high variance, and it is difficult to capture all user behavior with a
single model. Therefore, we segment users into clusters, and a model is
trained for each cluster. A subset of the training features are used
for clustering the users. Since most clustering algorithms suffer from
the curse of dimensionality, we want to avoid high-dimensional feature
vectors during clustering. 


Clustering is an important part of this framework, as it not only
groups users that should have similar data patterns but also does
filtering to some extent. In our experiments, we noticed that there is
one group that has a higher number of anomalies, which means anomalies
do share some level of similarities. As the number of normal data
points are still far larger than the anomalies, we are still able to
train quality models. For one of the data sets used in the experiments,
the classic k-means clustering is
chosen. Although k-means is a simple algorithm, it provides reasonably good results.
More sophisticated clustering algorithms might be
explored in future work. For the other data set, k-modes clustering
\cite{2001kmodes} is chosen. It is a k-means variant for categorical
data. 

To determine a proper $k$ for each algorithm, we use Silhouette
Coefficient (SC), which is defined as:
$SC(i) = \frac{b(i)-a(i)}{max\{ a(i), b(i) \}}$
, where $a(i)$ is the average dissimilarity to the intra-cluster data
points and $b(i)$ is the lowest average dissimilarity to a neighboring
cluster. For each data set, we try to find the $k$ with the highest average SC.  

\subsection{Baseline Methods}
Two baseline models—Principal Component Analysis (PCA) and Auto
Encoder (AE) are compared with the proposed method.
PCA and AE are general methods to learn low-dimensional embeddings.
These embeddings can be thought of as features learned from data. We
compare these with our proposed method, AS2S. A key difference with
AS2S is that each input vector is considered independent, while AS2S is
a sequence model and captures temporal relationships between subsequent
examples. 

\subsubsection{Principal Component Analysis}
PCA is a common statistical method to conduct dimension reduction for data. It uses orthogonal transformation to convert data into a new space that has linearly uncorrelated variables in each dimension. The
top $p$ principal components contain the most information of the data, where $p$ is a parameter of the model. A detailed description of PCA can be found in \cite{jolliffe2016principal}. 

\subsubsection{Auto Encoder}

The basic structure of AE is a simple 3-layer neural network that includes input, hidden, and output layers. The raw data is fed into the input layer, encoded into the hidden layer, and then decoded in
the output layer. The optimization aims at minimizing the difference between the input and the decoded output. To generate a dense representation for an input, we take the outputs of the hidden layer. The number of hidden units, $p$, is a parameter of the model that controls the size of the learned representations. The activation function used in our experiment is empirically chosen to be hyperbolic tangent (tanh). 

\subsection{Attention Sequence-to-Sequence Recurrent Neural Network Model}

In \cite{DBLP:journals/corr/MalhotraRAVAS16}, Malhotra et al. adopt Sequence-to-Sequence Recurrent Neural Network model (S2S) for anomaly detection in several time-series data, and show that S2S exhibits superior performance in detecting anomalies. The results inspired us to apply such a model to security data with temporal dependencies. \cite{2015neural,DBLP:journals/corr/HermannKGEKSB15,DBLP:journals/corr/XuBKCCSZB15} introduce a powerful new mechanism called “attention,” which provides the S2S model with the ability to find soft alignments between input and output sequences. This mechanism has shown promising results in numerous natural language tasks, such as machine translation, question answering, and artificial conversations (i.e., chat bots). However, this mechanism has not been previously applied to time-series and security data. The intuition that the soft-alignment can help to capture more accurate hidden state representations motivated us to experiment with AS2S. 

\begin{figure}
  \centering
  \includegraphics[width=0.8\textwidth]{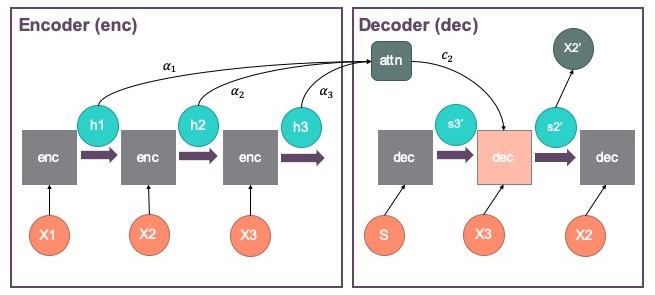}
  \caption{AS2S architecture showing how X2’ (second sequence element) is predicted}
    \label{fig:attn_rnn}
\end{figure}

Our model is most similar to \cite{2015neural}. Fig. \ref{fig:attn_rnn} shows the AS2S network architecture. The encoder and decoder are both Recurrent Neural Networks (RNN) with a Long Short Term Memory (LSTM) cell. The encoder network is defined as: 
\begin{equation}
h_t = f_{enc}(x_t, h_{t-1}),
\end{equation}
where $h_t$ is the hidden state and $x_t$ is the input at timestamp $t$; $f_{enc}$ is the LSTM cell, which could be multi-layered. At each timestamp the encoder RNN takes the previous hidden state $h_{i-1}$  and the current window $x_i$ as the inputs and outputs the current hidden state $h_i$. The decoder network is defined in a similar way but includes an attention component (network), which aims to capture the alignment between the input and output sequences. 
\begin{equation}
s_i = f_{dec}(x_{i+1}, s_{i+1}, c_i),
\end{equation}
where $x_i'$ is the $i$-th token in the output sequence, $s_i$ is the $i$-th hidden state, and $c_i$ is the context vector. Note that the sequence index is in the reversed order, e.g., from $x_k$ to $x_1$, which is a convention to empirically achieve better performance, i.e., the last token is easier to be decoded in early timestamps. Each context vector $c_i$ is computed by weighting the input hidden states.
\begin{equation}
c_i = \sum_{j=1}^{N} \alpha_{ij} h_j ,
\end{equation}
where $N$ is the sequence length, $i$ is the $i$-th timestamp of the decoder,
and $\alpha_{ij}$ is the weight for each hidden state $h_j$ from the encoder.
The weight $\alpha_{ij}$ is defined as
\begin{equation}
    \alpha_{ij} = \frac{exp(a_{ij})}{Z},
\end{equation}
\begin{equation}
    a_{ij} = attn(s_{i+1}, h_j),
\end{equation}
where $Z$ is the normalization term; $attn(.)$ is the attention network that takes the previous output state $s_{i+1}$ and $h_j$ as inputs, and outputs the logit capturing the alignment.  The attention component allows, at each timestamp, the network to consider the weighted encoder states to make inferences. Similar to the previous two models, AS2S also has a parameter $p$, which is the number of hidden units used in the encoder and decoder. To build the representations of each input window, we design the loss function to be squared errors between the input and the output, i.e., it is a sequence-to-sequence auto encoder with attention.

\section{Experimental Results} \label{sec:experiments}

To demonstrate the effectiveness of our method we use two real world
data sets, and compare our performance with the two selected
baseline models. 
To evaluate the models, we randomly sample 15\% of the users in each
cluster as the test set and 85\% as the training set. We train all the
models for 50 epochs. We use a development set from within the
training set for model selection. 
After features are learned, we use a simple anomaly detection algorithm based on $k$-nearest neighbors to compute an anomaly score. Then we vary the anomaly threshold to compute precision and recall. 
The performance of the models, based
on Area Under Precision-Recall Curves (PR-AUC) computed from partial
labels, on the anomaly detection task is used as a proxy for feature
learning.

\subsection{Data Set}

We refer to the two data sets as NETFLOW and CERT.

\subsubsection{NETFLOW}
The NETFLOW data set is from \cite{shiravi2012toward} and is converted into netflow format \cite{hofstede2014flow}, a commonly-used data source in the IT security industry. This data set spans 7 days and contains network-level information.
Three types of features — simple counts, bitmap, and top-K — are extracted. The features are directional. Half of the features are for incoming traffic and half for outgoing. For example, in count features, there are 60 features in total, 30 for outgoing traffic and the other 30 for incoming traffic. The bitmap features aggregate the flag bits used in the window.
For the top-k features, we empirically pick $k$ equal to 5. Note that since NETFLOW does not contain information like user profiles, we can only use a subset of the training features as the cluster features. In this case, count and bitmap are selected as the clustering features, since we recognize that they are more related to user behavior. In total, we identified 108 features for training. 
\subsubsection{CERT}
The CERT data set
\cite{2016insider} is quite large comparatively and has 516 days of
data with application-level information and user profiles. 
The features we use are similar to
\cite{tuor2017deep} but with some simplifications. The CERT data set
contains 
many application details and user profiles data as well. 
As a result, we can extract more features in this data set
than in the NETFLOW data set. For clustering, we use the categorical
information from user profiles, e.g., user roles, current projects,
teams, and so on, as the features, and the discrete indices as their
values. For training, 352 count features are selected.

\subsection{NETFLOW Results}

\begin{figure}
  \centering
  \includegraphics[width=0.45\textwidth, height=0.3\textwidth]{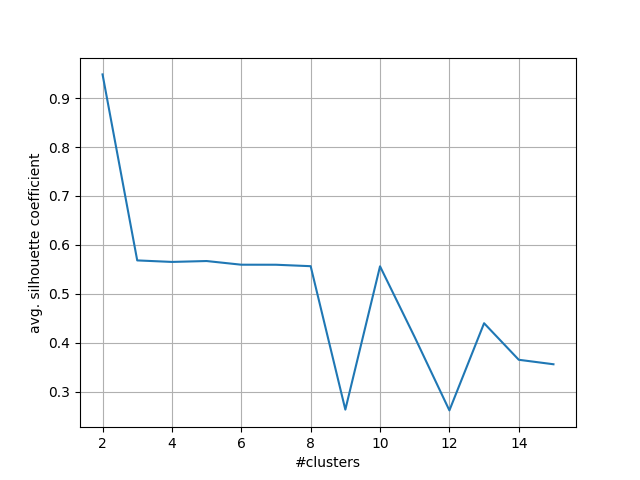}
  \caption{Determining the number of clusters K for NETFLOW}
    \label{fig:nclusters_netflow}
\end{figure}

\begin{figure}
  \centering
  \includegraphics[width=0.45\textwidth,  height=0.3\textwidth]{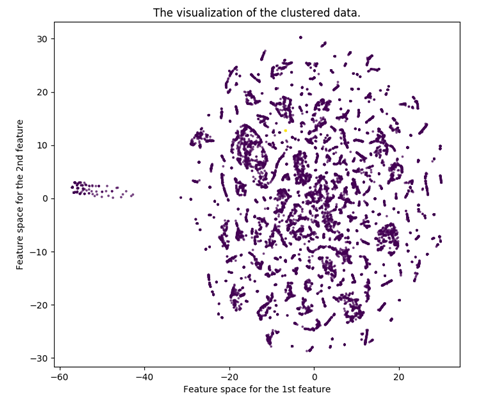}
  \caption{t-SNE visualization of NETFLOW}
    \label{fig:tsne_netflow}
\end{figure}

Fig. \ref{fig:nclusters_netflow} shows the quality of clustering as
measured by the Silhouette Coefficient (SC) with respect to the number
of clusters, $k$ of k-means. We can see that $k=2$ gives us the best SC
score. We can get an intuitive validation of this clustering result by
looking at Fig. \ref{fig:tsne_netflow}, the t-SNE visualization of all
the data points. t-SNE usually provides a good
visualization of high-dimension data; this visualization shows a clear
margin between the two clusters and that one cluster is far
larger than the other. If we consider the labels of the data points,
we observe the smaller cluster (on left, called cluster 1) has a higher
proportion of anomalies (of 161 points, 35 are anomalous), which implies that clustering successfully
filters the anomalies to some extent. For cluster 0 (on right), it has an
extremely small number of anomalies (of 100 K points, 36 are anomalous), which makes the detection much harder. All methods perform poorly in detecting anomalies in cluster 0.


\begin{figure}[]
        \centering
        \includegraphics[width=0.45\textwidth,height=0.3\textwidth]{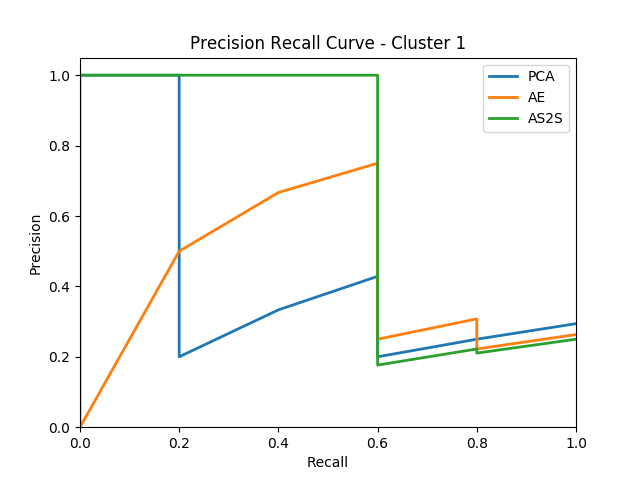}
    \caption{Anomaly detection results on the NETFLOW data set}
    \label{fig:ad_netflow}
\end{figure}

Fig. \ref{fig:ad_netflow} presents the anomaly detection results for
cluster 1.
Up to recall of 60\%, AS2S provides higher precision, while at higher recall all the three methods perform similarly. 
Overall AS2S outperforms the other two baselines based on AUC,
showing that temporal relationship and attention
mechanism help anomaly detection in this scenario. 

\subsection{CERT Results}

\begin{figure}
  \centering
  \includegraphics[width=0.45\textwidth, height=0.3\textwidth]{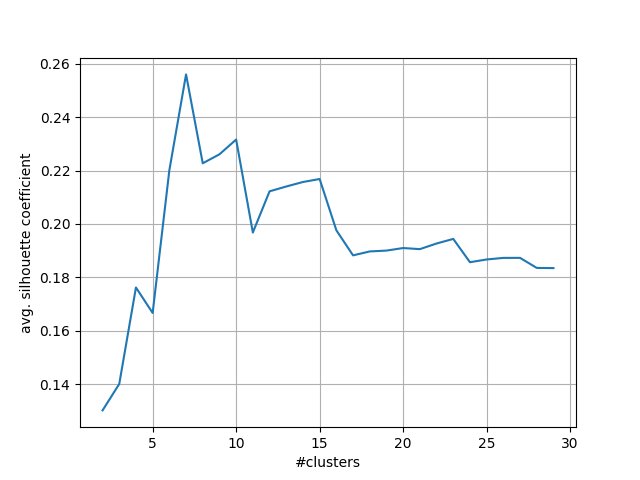}
  \caption{Determining the number of clusters K for the CERT data set}
    \label{fig:nclusters_cert}
\end{figure}


Fig. \ref{fig:nclusters_cert} shows the quality of clusters as
measured by the SC score with respect to the number of clusters,
$k$. $k=7$ is selected as it gives us the best SC
score.
The number of data points in the clusters are
distributed quite uniformly. Only one cluster is relatively a bit
larger.

\begin{figure}
  \centering
  \includegraphics[width=0.45\textwidth, height=0.3\textwidth]{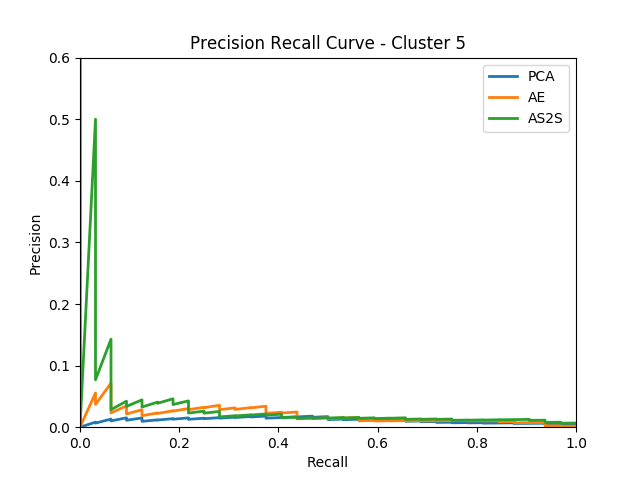}
  \caption{Anomaly detection results for cluster 5 of CERT.}
    \label{fig:ad_cert}
\end{figure}

In the CERT data set, there are only 47 anomalies out of nearly 1M data
points. When we delve into the distribution of these anomalies, we 
observe that only 3 clusters (cluster 0, 5, and 6) have anomalies, and
only cluster 5 has enough anomalies (32 anomalies) to provide
statistically meaningful results. Therefore, in
Fig. \ref{fig:ad_cert}, we only show the results for cluster 5. AS2S
has the highest area under the precision-recall curve, followed by AE
and then PCA. The results show that, similar to the NETFLOW
results, AS2S is able to detect anomalies better than the
baselines.
\vspace{-0.5em}
\subsection{Discussion}


Feature learning in images and natural languages is obvious for
humans. Learned image features like texture, edges, objects require no
explanations. However, the same is not true for security data. Here we
use performance on an anomaly detection task as a proxy for
quality of feature learning, quantified by PR-AUC. As the AS2S
model was trained, we wanted to explore if the features learned 
improved. Note that PR-AUC
already captures this quantitatively for all the final models.     

\begin{figure}[htb]
    \centering
    \begin{subfigure}
        \centering
        \includegraphics[width=0.42\textwidth]{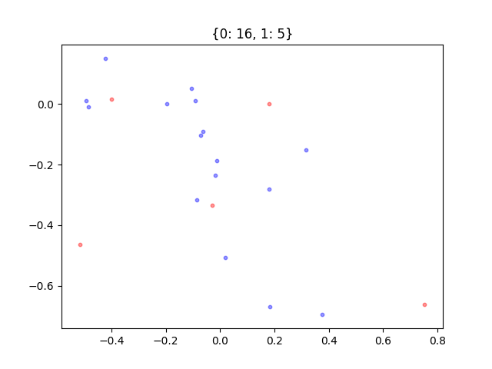}
    \end{subfigure}%
    ~ 
    \begin{subfigure}
        \centering
        \includegraphics[width=0.42\textwidth]{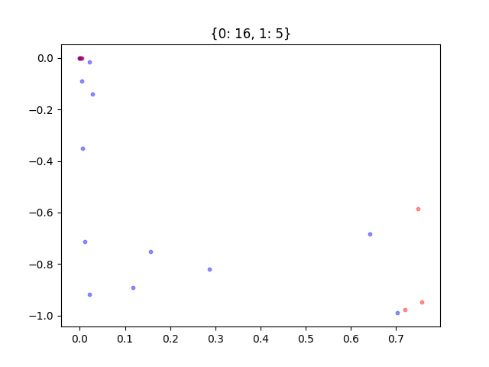}
    \end{subfigure}
    ~ 
    \begin{subfigure}
        \centering
        \includegraphics[width=0.42\textwidth]{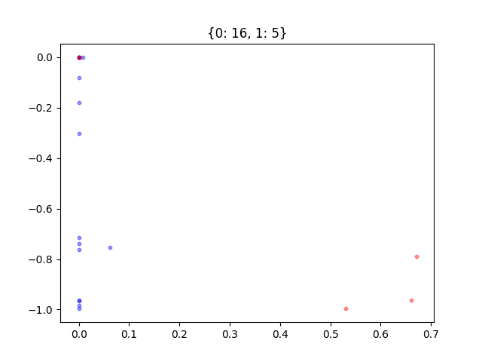}
    \end{subfigure}
    \caption{Scatter plots of features learned with AS2S model from the test set of cluster 1 in NETFLOW. From top to bottom, the plot is from epoch 1, 15, and 24 respectively. The anomalies are in red and the normal data points are in blue.}
    \label{fig:scatter_netflow}
\end{figure}

Fig. \ref{fig:scatter_netflow} shows the scatter plots of the hidden
layer outputs for AS2S projected onto two dimensions.
It is from  
cluster 1 in the NETFLOW test data set during training.
Out of 21 points, 5 are anomalous in this test set,
so we can clearly see the changes during training. In the early epoch
(top), all the data points distribute randomly. In epoch 15 (middle)
some points aggregate to the left and some to the right. In epoch 24
(bottom), we can see that 3 anomalies aggregate to the right, which is
good. However, some other anomalies are on the left together with the
normal points. This means that, on one hand, some anomalies have clear
properties and can be identified by AS2S; on the other hand, some are
similar to the normal nodes, and therefore are unlikely to be
detected. These plots demonstrate the effectiveness of AS2S to
progressively learn more meaningful representations of NETFLOW for
detecting anomalies without any labels.

We use a precision-recall curve instead of a ROC curve, since the data is
highly skewed (as expected because anomalies are rare), and ROC curves
are misleading is such cases. Furthermore, precision, which is the
fraction of true positives (out of all positives marked by a model),
directly quantifies the overhead on a security operations center
analyst. While ideally one would prefer both high precision and high recall,
for a particular classifier higher precision is usually preferred to 
prevent alarm fatigue (and since in practice a large number of
classifier are simultaneously deployed, the recall across the entire
system may still be good).

\vspace{-0.5em}
\section{Conclusions}
We explored multiple methods for unsupervised and self-supervised
feature learning for security data. To the best of our knowledge, this
is the first attempt to apply the attention mechanism (AS2S) in
security anomaly detection. We demonstrated that AS2S generally performs well
for anomaly detection, especially for NETFLOW data, outperforming PCA
and AE.
However, anomaly
detection for clusters with extremely small number of anomalies is
difficult for all of the models we experimented with.
These factors reveal the difficulties of unsupervised and self-supervised
feature learning from security data and  further research is required to
address them.



\bibliographystyle{unsrt}  
\bibliography{references}

\end{document}